\documentclass[%
reprint,
notitlepage,
amsmath,amssymb,amsthm,
aps,
pra
]{revtex4-1}

\usepackage{xcolor}
\usepackage{graphicx}
\usepackage{dcolumn}
\usepackage{bm}
\usepackage[hidelinks]{hyperref}

\newcommand{\ket}[1]{\lvert #1\rangle}

\newcommand{\braket}[2]{\langle #1\vert#2\rangle}
\newcommand{\abs}[1]{\lvert #1\rvert}
\newcommand{\ii}{\mathrm{i}}

\newtheorem{theorem}{Theorem}
\newtheorem{lemma}{Lemma}

\begin{document}

\title{Geometric Floquet Condition for Quantum Adiabaticity}

\author{Jie Gu}
 \email{jiegu1989@gmail.com}
 \affiliation{Chengdu Academy of Education Sciences, Chengdu 610036, China}
 
\author{X.-G. Zhang}%
\email{xgz@ufl.edu}
\affiliation{%
 Department of Physics, the Quantum Theory Project and the Center for Molecular Magnetic Quantum Materials, University of Florida, Gainesville 32611, USA
}%
\affiliation{Quantum Potential PTE. LTD., Singapore 149739}

\date{\today}

\begin{abstract}
Quantum adiabaticity is the evolution of a quantum system that remains close to an instantaneous eigenstate of a time-dependent Hamiltonian.
Using Floquet formalism, we derive a rigorous sufficient condition for adiabaticity in closed, finite-dimensional periodically driven systems that is valid for arbitrarily many driving periods.
The condition is stroboscopic and geometric, depending only on single-cycle information: the Fubini--Study length of the instantaneous eigenray and a quasienergy-separation measure extracted from the Floquet operator.
We also formulate a state-targeted refinement that reduces conservativeness when only one adiabatic branch is relevant.
Rather than synthesizing control pulses, the result provides a certification criterion for a given periodic protocol.
We illustrate the criterion and contrast it with conventional instantaneous-gap conditions in three representative examples.
\end{abstract}

\maketitle


\section{Introduction}
The quantum adiabatic theorem (QAT) \cite{ehrenfest1916,born1928} is a cornerstone of near-equilibrium quantum dynamics, underpinning results and applications ranging from the Gell-Mann--Low theorem \cite{PhysRev.84.350} to adiabatic quantum computation \cite{doi:10.1126/science.1057726}.
In its usual form, QAT supports the intuition that adiabatic evolution requires ``slow'' driving, so that the system changes quasi-statically.
This intuition becomes ambiguous in periodically driven (Floquet) systems, which are ubiquitous in applications and continue to attract sustained theoretical and experimental attention (see, e.g., \cite{grifoni1998,bukov2015universal,eckardt2017,oka2019floquet} for reviews).
Because Floquet dynamics over one period is encoded in a time-independent stroboscopic description, it is natural to ask whether adiabatic following can also be formulated---and guaranteed---in a way that explicitly exploits time periodicity.

We adopt the standard definition of quantum adiabaticity (QA) \cite{Comparat2009,Comparat2011}:
for $t\in[t_0,t_0+\tau]$, the state $\ket{\Psi(t)}$ evolving from an instantaneous eigenstate $\ket{E_n(t_0)}$ remains close to that eigenstate in the sense that the overlap
$\abs{\braket{E_n(t)}{\Psi(t)}}$ stays close to unity.
Motivated by static systems, much of the literature quantifies adiabaticity through the time variation of instantaneous eigenstates, leading to the traditional criterion (we set $\hbar=1$)
\begin{equation}
\label{eq:cond_tr}
\max_{m\neq n} \bigl| { \langle E_m(t)|\dot E_n(t) \rangle} \bigr| \ll \min_{m\neq n} \left|{E_m(t)-E_n(t)} \right| ,
\end{equation}
with the maximum and minimum taken over $t\in[t_0,t_0+\tau]$.
However, Eq.~\eqref{eq:cond_tr} is neither sufficient nor necessary \cite{Marzlin2004,Tong2005,Du2008}.
Notably, most counterexamples highlighting the failure of Eq.~\eqref{eq:cond_tr} involve periodically driven systems, where resonance-like oscillations can induce transitions even when the instantaneous gap is large \cite{Marzlin2004,Tong2005,wu2005validity,MacKenzie2007,Amin2009,Comparat2009,ortigoso2012}.

Several resolutions have been proposed \cite{tong2007,Wu2008,Comparat2009,wang2016}, but they do not explicitly leverage time periodicity.
While these criteria are sufficient, many practical forms become restrictive or cumbersome in high-frequency driven systems.
Moreover, because their derivations are typically formulated for evolution over a finite duration, they do not directly address a basic practical question:
given a system with known energy structure, what class of {time-periodic} drives (e.g., microwave control) will {never} excite it, even after indefinitely many cycles?

A related viewpoint, emphasized qualitatively by Russomanno and Santoro \cite{russomanno2017}, is that adiabaticity in Floquet systems should require the driving frequency to remain sufficiently far from quasienergy degeneracies.
However, a quantitative, stand-alone condition that incorporates this ``farness'' and guarantees QA uniformly in time has been missing.
In this work, we derive a simple and fully rigorous sufficient criterion for QA in periodically driven systems within the Floquet formalism.
Our geometric Floquet condition [Eq.~\eqref{eq:cond_main}] is stroboscopic and geometric: it depends on the one-cycle Fubini--Study length traversed by the instantaneous eigenray and a quasienergy-separation measure that quantifies the distance from quasienergy degeneracies modulo the drive frequency.
Crucially, the resulting bound is {uniform in time}: whenever the geometric Floquet condition is satisfied, the fidelity with the chosen instantaneous eigenstate remains above a prescribed threshold for all $t\ge t_0$, i.e., for arbitrarily many driving periods.
Although written in terms of quasienergies, the separation measure depends only on the spectrum of the one-period evolution operator (Floquet operator) and can be evaluated without explicitly unfolding quasienergies; see Eq.~\eqref{eq:operator-spectrum}.
A useful state-targeted refinement further reduces conservativeness when only a designated adiabatic branch matters.
The theorem applies to closed, finite-dimensional, exactly periodic Hermitian Hamiltonians with a smoothly chosen nondegenerate instantaneous eigenray; robustness to noise, dissipation, or slow envelope modulation is left for future work.
We illustrate the power of the geometric Floquet condition and contrast it against conventional criteria using three representative examples.


\section{Setup and main results}
Let $H(t)=H(t+T)$ be a Hermitian Hamiltonian on an $N$-dimensional Hilbert space with the drive angular frequency $\omega \equiv 2\pi/T$, and let
$\ket{E_n(t)}$ be a normalized instantaneous eigenstate whose eigenvalue remains isolated and nondegenerate over one period.
We assume that the corresponding eigenray is chosen smoothly in $t$.
Because $H(t_0)=H(t_0+T)$ and the eigenvalue is nondegenerate, the {ray} of $\ket{E_n(t)}$ is $T$-periodic; we will assume a periodic gauge,
\begin{equation}
\ket{E_n(t+T)}=\ket{E_n(t)},
\label{eq:periodic_gauge}
\end{equation}
which can always be enforced by a smooth phase choice and does not affect fidelities \cite{Berry1984}.
Fix an initial time $t_0$ and let $\ket{\Psi(t)}$ solve
\[
\ii\,\partial_t\ket{\Psi(t)}=H(t)\ket{\Psi(t)},\qquad \ket{\Psi(t_0)}=\ket{E_n(t_0)}.
\]
Define the instantaneous overlap
\begin{equation}
d_n(t)\equiv \braket{E_n(t)}{\Psi(t)}.
\label{eq:def_dn}
\end{equation}

Let $\{\ket{\phi_\alpha(t)},\epsilon_\alpha\}$ be a complete Floquet set \cite{Shirley1965,Sambe1973}:
\[
\bigl[H(t)-\ii\partial_t\bigr]\ket{\phi_\alpha(t)}=\epsilon_\alpha\ket{\phi_\alpha(t)},
\qquad \ket{\phi_\alpha(t+T)}=\ket{\phi_\alpha(t)}.
\]

Our main result is the following theorem and its corollary.
\begin{theorem}
\label{thm:main}
Assume $g>0$ and $\mathcal{L}_n \le {\pi}/{2}$.
Then the loss of fidelity amplitude, $1-|d_n(t)|$, is bounded uniformly in time as
\begin{equation}
\label{eq:bound_main}
1-|d_n(t)| \le \left[\frac{2\sin(\mathcal{L}_n/2)}{g}\right]^2,
\end{equation}
where the one-period Fubini--Study length \cite{ProvostVallee1980,AnandanAharonov1990}
\[
\mathcal{L}_n \equiv \int_{t_0}^{t_0+T} v_n(t)\, \mathrm{d}t
\]
uses the gauge-invariant Fubini--Study speed
\begin{equation}
\begin{aligned}
v_n(t) &\equiv
\sqrt{\langle \dot{E}_n(t)|\dot{E}_n(t)\rangle-|\langle E_n(t)|\dot{E}_n(t)\rangle|^2} \\
&=\sqrt{\sum_{m\neq n}\abs{\langle E_m(t)|\dot E_n(t)\rangle}^2},
\end{aligned}
\label{eq:FS_speed}
\end{equation}
and
the (mod-$\omega$) quasienergy-separation measure
\begin{equation}
\label{eq:g_def}
g \equiv \min_{\alpha\neq \beta}
\left|
\sin\!\left(\frac{\pi(\epsilon_\alpha-\epsilon_\beta)}{\omega}\right)
\right| .
\end{equation}
\end{theorem}
The proof is given in the Appendix.

A straightforward corollary is as follows: For any prescribed accuracy $0<\varepsilon \ll 1$, the loss of fidelity amplitude is upper bounded uniformly in time by $\varepsilon$ if
\begin{equation}
\label{eq:cond_main}
\frac{\mathcal{L}_n}{g} \le \sqrt{\varepsilon}.
\end{equation}
We refer to Eq.~\eqref{eq:cond_main} as the geometric Floquet condition.
No assumption such as $\omega \ll \min_{m\neq n}|E_m-E_n|$ is imposed.

When the focus is a designated adiabatic branch rather than the worst-case Floquet spectrum, one may sharpen the criterion as follows.
Let $\alpha_0$ denote the Floquet label with the largest overlap with $\ket{E_n(t_0)}$ and define the targeted stroboscopic gap
\begin{equation}
\label{eq:g_targeted_def}
g_{\alpha_0}\equiv \min_{\beta\neq\alpha_0}
\left|
\sin\!\left(\frac{\pi(\epsilon_{\alpha_0}-\epsilon_\beta)}{\omega}\right)
\right| .
\end{equation}
The argument in the Appendix then yields the same uniform-in-time estimate as Eq.~\eqref{eq:bound_main} with $g$ replaced by $g_{\alpha_0}$.
This state-targeted variant is never weaker than the global criterion and can be substantially less conservative when near-degeneracies among spectator Floquet branches are irrelevant to the selected instantaneous eigenstate.


\section{Remarks and connection with existing conditions}
Two remarks are immediate.
First, the geometric Floquet condition is derived rigorously, without uncontrolled approximations.
Second, the bound is {uniform in time}: if Eq.~\eqref{eq:cond_main} holds, the instantaneous-eigenstate fidelity stays close to unity for arbitrarily long evolution, i.e., for an unlimited number of driving periods.

Theorem~\ref{thm:main} also clarifies how Eq.~\eqref{eq:cond_main} sits relative to standard adiabatic criteria.
For general $H(t)$, rigorous adiabatic theorems are most naturally formulated in terms of instantaneous eigenprojectors (beginning with Kato) and yield finite-time error bounds in terms of $\dot H(t)$-type norms and inverse powers of an instantaneous spectral gap \cite{Kato1950,AvronElgart1999,Jansen2007}.
In a strictly periodic setting, however, one often asks whether a drive can be repeated indefinitely without transitions accumulating.
Theorem~\ref{thm:main} exploits Floquet structure: once the state is dominated by a single Floquet mode, its occupation is exactly conserved, so adiabaticity does not degrade with the number of cycles.
This is why a one-cycle condition can imply the time-uniform bound~\eqref{eq:bound_main}.

Structurally, Eq.~\eqref{eq:cond_main} mirrors the textbook instantaneous-gap inequality~\eqref{eq:cond_tr}.
The numerator is encoded by the Fubini--Study speed $v_n(t)$ and the one-cycle length $\mathcal{L}_n=\int_{t_0}^{t_0+T} v_n(t)\,\mathrm{d}t$, a gauge-invariant measure of instantaneous eigenstate rotation.
The essential change is in the ``denominator'': the instantaneous energy gap is replaced by the quasienergy-separation measure $g$, i.e., the distance to any quasienergy degeneracy modulo $\omega$ (multiphoton resonances).
This replacement resolves periodically driven counterexamples to Eq.~\eqref{eq:cond_tr}: even with large instantaneous gaps, transitions can be enabled by resonant quasienergy structure \cite{Marzlin2004,Tong2005,wu2005validity,Comparat2009,ortigoso2012}.
Accordingly, Eq.~\eqref{eq:cond_main} makes quantitative the expectation articulated qualitatively in Ref.~\cite{russomanno2017} that adiabaticity requires driving sufficiently far from quasienergy degeneracies.

A large literature refines Eq.~\eqref{eq:cond_tr} using integral equations and more careful phase bookkeeping \cite{tong2007,Wu2008,Amin2009,Comparat2009,wang2016}.
These results are typically finite-duration statements, and many practical forms become conservative in large Hilbert spaces by bounding $O(N^2)$ sums of couplings via maxima (see, e.g., Eq.~(16) of Ref.~\cite{tong2007}, Eq.~(25) of Ref.~\cite{Amin2009}, Eq.~(19) of Ref.~\cite{Wu2008}, and Eq.~(16) of Ref.~\cite{wang2016}).
By contrast, Eq.~\eqref{eq:FS_speed} packages all instantaneous couplings into the single gauge-invariant norm $v_n(t)$, and Theorem~\ref{thm:main} yields a stroboscopic one-cycle criterion with no explicit $N$-prefactor.
The state-targeted refinement based on $g_{\alpha_0}$ sharpens this further when only one selected adiabatic branch matters and near-degeneracies among spectator Floquet modes are irrelevant.
Eq.~\eqref{eq:cond_main} can thus be viewed as a periodic-drive-specialized sufficient condition that complements general integral formulations by trading detailed time-history control for a purely one-cycle control parameter.

It is also important to distinguish the present result from control-synthesis approaches.
Shortcuts to adiabaticity and invariant-based engineering aim at exact finite-time state transfer by modifying the Hamiltonian, whereas optimal-control methods numerically design pulses that maximize a task-specific cost function on a finite time horizon \cite{Torrontegui2013,GueryOdelin2019,Khaneja2005,Caneva2011,Reich2012}.
Our theorem does not attempt to replace or outperform those methods.
Instead, it supplies an analytic {certificate} for a fixed implemented periodic drive: if Eq.~\eqref{eq:cond_main} or its state-targeted refinement is satisfied, repeated use of that drive cannot accumulate arbitrary excitation.
This niche is especially relevant when the objective is to preserve the instantaneous eigenstate of the available Hamiltonian rather than to realize an alternative engineered path.

Finally, our setting should be distinguished from the adiabatic Floquet principle, where one prepares a Floquet state for a fixed drive and varies an external parameter slowly compared to $T$ (e.g., a slow envelope ramp) \cite{BreuerHolthaus1989,grifoni1998}.
There adiabaticity is formulated in the extended Sambe space and hinges on quasienergy gaps \cite{Shirley1965,Sambe1973}; because quasienergies are defined modulo $\omega$, the relevant issue becomes whether near-resonant avoided crossings are ineffective \cite{grifoni1998}.
The proliferation of weak avoided crossings and the possible absence of a strict Floquet adiabatic limit in large Hilbert spaces were emphasized in Ref.~\cite{Hone1997} and analyzed further in Ref.~\cite{Weinberg2017}.
Theorem~\ref{thm:main} addresses a complementary question: for a {fixed} periodic drive, when is an instantaneous {energy} eigenstate an adiabatic invariant for arbitrarily many cycles?
The answer is again controlled by quasienergy resonances, through the single-cycle mixing encoded by $\mathcal{L}_n$ together with the stroboscopic gap $g$ (or $g_{\alpha_0}$ in the state-targeted version).
In particular, multiphoton avoided crossings appear as small-gap regions, where an instantaneous eigenstate necessarily has appreciable weight on multiple Floquet modes and long-time beating becomes unavoidable unless additional structure suppresses the mixing, consistent with extended-space perturbative treatments \cite{RodriguezVega2018}.
This restriction to exact periodicity is the price paid for the time-uniform certification.


\section{Experimental considerations}
The geometric Floquet condition is formulated in terms of quantities accessible from one driving cycle, at least in principle.
The quasienergy-separation measure $g$ can be obtained from the eigenvalues \(\{\lambda_\alpha\}\) of the one-period propagator $U(t_0+T,t_0)=\mathcal{T}\exp\!\left[-\ii\int_{t_0}^{t_0+T}H(t)\, \mathrm{d}t \right]$ (with \(\lambda_\alpha=e^{-i\epsilon_\alpha T}\) and \(T=2\pi/\omega\)):
\begin{equation}
\label{eq:operator-spectrum}
\min_{\alpha\neq\beta}\left|\sin\!\left(\frac{\pi(\epsilon_\alpha-\epsilon_\beta)}{\omega}\right)\right|
=\frac12\min_{\alpha\neq\beta}\abs{\lambda_\alpha-\lambda_\beta},
\end{equation}
Thus $g$ is half the minimum chord distance between Floquet eigenvalues on the unit circle.
For few-level platforms one may reconstruct $U(t_0+T,t_0)$ by applying the drive for one period to a set of input states and performing state or process tomography.
For larger Hilbert spaces such full tomography becomes expensive, so our accessibility claim should be interpreted most directly for low-dimensional, symmetry-reduced, or otherwise controllable sectors, or for platforms where the eigenphases of $U(t_0+T,t_0)$ can be extracted interferometrically or via phase-estimation protocols.
In all cases, only the spectrum of $U(t_0+T,t_0)$ is needed; explicit quasienergy branch unfolding is unnecessary.
The geometric factor $\mathcal{L}_n$ depends only on the instantaneous eigenstates of the calibrated Hamiltonian and can be computed from the control waveform $H(t)$.
Therefore the criterion functions primarily as a single-cycle diagnostic and certification tool for a specified periodic protocol.
A quantitative analysis of finite-sampling error, Hamiltonian miscalibration, and decoherence in the inferred bound is beyond the present work and remains important for extending the criterion to noisy experiments.


\section{Examples}
\subsection{Example 1: Schwinger-Rabi model}
Consider a two-level system driven by an oscillatory field \cite{schwinger1937, rabi1937},
\begin{equation}
\label{eq:ham}
H(t)
= \frac{\omega_0}{2}\begin{pmatrix}
\cos \theta & \sin \theta\, e^{-i\omega t} \\
\sin \theta\, e^{i\omega t} & -\cos \theta
\end{pmatrix},
\end{equation}
which is exactly solvable and has been extensively studied as a benchmark model for adiabatic conditions
\cite{Tong2005, Comparat2009, Tong2010}.
Here, $\omega_0$, $\theta$, and $\omega$ are constants, with $\theta \in [0,\pi)$.
The instantaneous energy eigenvalues are $E_{0,1}=\mp \omega_0/2$, with corresponding eigenstates
\[
|E_{0}\rangle  =
\begin{pmatrix}
e^{-i\omega t/2 } \sin \frac{\theta}{2} \\
- e^{i\omega t/2} \cos \frac{\theta}{2}
\end{pmatrix},
\,\,\,
|E_{1}\rangle  =
\begin{pmatrix}
e^{-i\omega t/2 } \cos \frac{\theta}{2} \\
e^{i\omega t/2} \sin \frac{\theta}{2}
\end{pmatrix}.
\]
Assuming the system is prepared in the ground state at $t=0$, the overlap between the evolved state and the
instantaneous ground state is
\[
\label{eq:overlap1}
|d_0(t)| = \sqrt{1-\left( \frac{\omega \sin \theta}{\Omega} \sin \frac{\Omega t}{2} \right)^2},
\]where
$\Omega = \sqrt{\omega_0^2+\omega^2 - 2\omega \omega_0 \cos \theta}$.

For evolution over arbitrarily long times, the sufficient and necessary  condition for QA in this model is
$1-|d_0(t)|\ll 1$, which reduces to
\begin{equation}
\label{eq:cond_1}
\frac{1}{2}\left(\frac{\omega \sin \theta}{\Omega} \right)^2 \ll 1.
\end{equation}
Equation~\eqref{eq:cond_1} can be satisfied in three parameter regimes:

(1) $\sin \theta \ll 1$ with $\theta > \pi/2$, while $\omega$ can be arbitrary;

(2) $\omega \ll \omega_0$, while $\theta$ can be arbitrary;

(3) $\sin \theta \ll 1$ with $\theta < \pi/2$, and $\omega$ must be detuned from resonance.

By contrast, the traditional condition Eq.~\eqref{eq:cond_tr} yields
$\frac{1}{2}\omega \sin \theta \ll \omega_0$.
It is straightforward to construct two counterexamples:

{Counterexample 1:} $\sin \theta \ll 1$, $\theta<\pi/2$, and $\omega=\omega_0$ satisfies the traditional condition but violates Eq.~\eqref{eq:cond_1}, showing that the traditional condition is not sufficient.

{Counterexample 2:} $\sin \theta \ll 1$, $\theta>\pi/2$, and $\omega \gg \omega_0$ satisfies Eq.~\eqref{eq:cond_1} but violates the traditional condition, showing that the traditional condition is not necessary.

We next verify the geometric Floquet condition.
For this system, the Floquet quasienergies are $\epsilon_{0,1}=\pm \Omega/2$, and $\mathcal{L}_0=\pi|\sin\theta|$.
Therefore the geometric Floquet condition Eq.~\eqref{eq:cond_main} becomes
\[
\label{eq:cond_1f}
\frac{\pi |\sin \theta|}{|\sin ({\pi \Omega}/{\omega})|} \ll 1.
\]
Using the elementary bound $|\sin x|\le |x|$ immediately implies Eq.~\eqref{eq:cond_1}.
This confirms that Eq.~\eqref{eq:cond_main} is sufficient for QA.
In this two-level setting the state-targeted refinement coincides with the global criterion, because there are no spectator Floquet branches.
Moreover, because the geometric Floquet condition enforces $|\sin\theta|\ll 1$, it covers only Cases (1) and (3).

\begin{figure}[htb!]
\centering
\includegraphics[width=\linewidth]{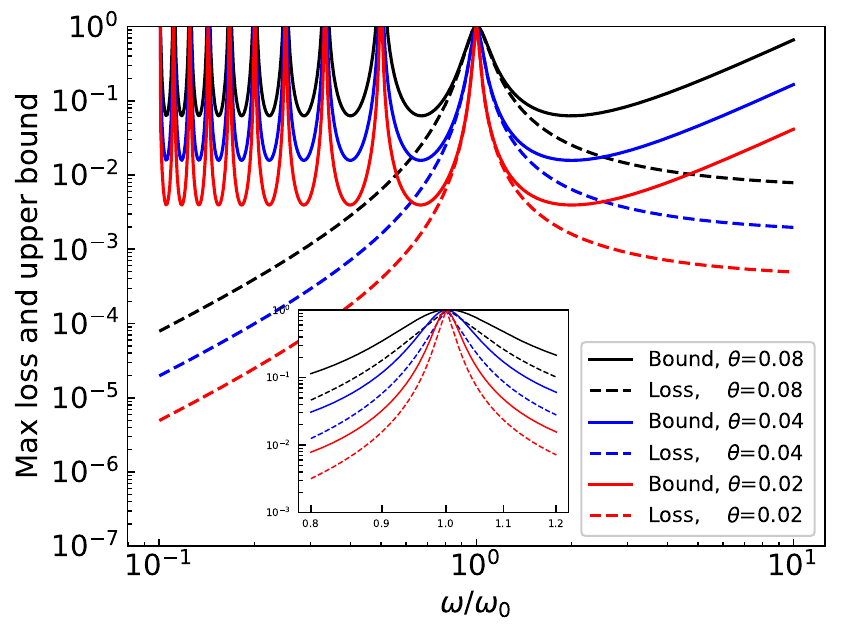}
\caption{\label{fig:f1}
Maximum loss of fidelity amplitude  (dashed lines) and the upper bound (solid lines) as a function of $\omega/\omega_0$ for different $\theta$ values (different colors).
The inset shows $\omega/\omega_0 \in [0.8,1.2]$.
}
\end{figure}

Equation~\eqref{eq:bound_main} is verified in Fig.~\ref{fig:f1}.
When $\theta \ll 1$, the loss of fidelity amplitude remains small except near $\omega \approx \omega_0$, and it stays below the quasienergy-gap bound at all times.
The bound diverges at quasienergy degeneracies (Floquet resonances), which occur at
\[
\omega_k = \left \{
\begin{aligned}
&\frac{\omega_0}{2\cos \theta}, \quad k=1\\
&\frac{\sqrt{\cos^2 \theta +(k^2-1)} - \cos \theta}{k^2-1} \omega_0 , \quad k\ge 2
\end{aligned}
\right .
\]
The quasienergy gaps become denser as $k \to \infty$ or $\omega_k/\omega_0 \to 0$ \citep{russomanno2017}.
At $\omega_k$ the geometric Floquet condition cannot be satisfied, yet QA can still hold.
In other words, Floquet resonance does not necessarily imply a breakdown of adiabaticity \cite{RodriguezVega2018}.
In the neighborhood of $\omega_0$, there is no quasienergy degeneracy; nevertheless, both the loss and the bound exhibit a pronounced peak.
This peak is excluded by the geometric Floquet condition, which is crucial for guaranteeing adiabaticity.


\subsection{Example 2: Dual hamiltonian}
As a second example, we consider the dual Hamiltonian of the model studied above, as discussed in Refs.~\cite{Tong2005, Duki2005,MacKenzie2007,ortigoso2012,Guo2013}. It is defined by
\[
\overline{H}(t) = -U^\dagger(t)\,H(t)\,U(t),
\]
where \(H(t)\) is given in Eq.~\eqref{eq:ham} and \(U(t)\) is the time-evolution operator generated by \(H(t)\). Throughout, an overbar denotes a quantity associated with the dual system. The dual Hamiltonian \(\overline{H}(t)\) is also periodic~\cite{Tong2005,Guo2013}, with period \(\overline{T}=2\pi/\Omega\).

The overlap between the state evolved from the initial ground state and the instantaneous ground state reads \footnote{The right-hand side of Eq.~(32) in Ref.~\cite{Tong2005} should read $\sqrt{1-\sin^2\theta \sin^2(\omega t/2)}$.}
\[
|d_0(t)|=\sqrt{1-\sin^2\theta\,\sin^2(\omega t/2)} .
\]

For long times, the sufficient and necessary condition for QA is
\begin{equation}
\label{eq:cond_2}
\frac{\sin^2\theta}{2}\ll 1.
\end{equation}
Diagonalizing the one-period propagator \(\overline{U}(\overline{T})=U^\dagger(\overline{T})\)~\cite{Tong2005,Guo2013}, we obtain quasienergies \(\pm \omega/2\) (modulo \(\Omega\)). The geometric Floquet condition in Eq.~\eqref{eq:cond_main} then yields
\begin{equation}
\label{eq:cond_2f}
\frac{\frac{\pi\omega|\sin\theta|}{\Omega}}{\left|\sin\!\left(\frac{\pi\omega}{\Omega}\right)\right|}\ll 1 .
\end{equation}
Using the inequality \( |\sin x|\le |x| \) reduces Eq.~\eqref{eq:cond_2f} to Eq.~\eqref{eq:cond_2}. By contrast, the traditional condition \(\omega\sin\theta\ll\omega_0\) does not reproduce Eq.~\eqref{eq:cond_2}.

\begin{figure}[htb!]
\includegraphics[width=\linewidth]{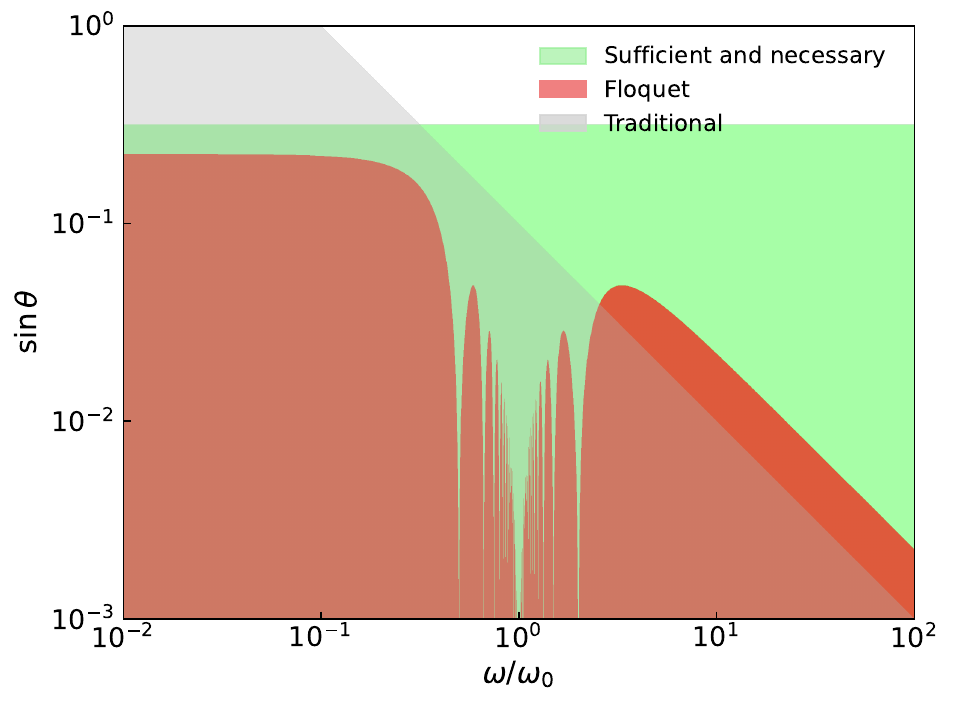}
\caption{\label{fig:f2}
Parameter regions in which each criterion ensures that the maximum loss of fidelity amplitude does not exceed \(\varepsilon=0.05\).}
\end{figure}

This analysis is corroborated in Fig.~\ref{fig:f2}, which compares the parameter regions predicted by each condition. For concreteness, we set the prescribed accuracy to \(\varepsilon=0.05\). The sufficient and necessary condition for QA then becomes \(|\sin\theta|\le 0.31\), while the geometric Floquet condition becomes
$
{\frac{\pi \omega |\sin\theta|}{\Omega}}/{\left|\sin\!\left(\frac{\pi\omega}{\Omega}\right)\right|}\le 0.22
$.
The gray region corresponds to the sufficient and necessary condition, and the light-green region shows the traditional criterion. Note that the traditional condition (and many related adiabatic criteria) is typically stated only up to the ambiguous symbol ``\(\ll\)'', which complicates practical use. To make it quantitative in the plot, we take it to mean
$\max_{m\neq n} \bigl| { \langle E_m(t)|\dot E_n(t) \rangle} \bigr| / \min_{m\neq n} \left|{E_m(t)-E_n(t)} \right| \le 0.05$.
Choosing a threshold other than \(0.05\) does not change our qualitative conclusion: the traditional criterion is neither sufficient nor necessary as can be seen from the plot. The geometric Floquet condition is shown in red and is entirely contained within the gray region. Moreover, it yields a broad admissible parameter range, and is particularly informative in the low-frequency regime, where it remains fairly sharp.


\subsection{Example 3: Many-body collective Ising model}
To demonstrate that the geometric Floquet condition remains predictive beyond two-level models---and, in particular, that it mitigates the scaling issue with Hilbert-space dimension that often arises in many-state adiabatic criteria---we consider an interacting system of $N_s$ spins-$\tfrac12$ with collective infinite-range Ising interactions subject to a periodic transverse drive \cite{LipkinMeshkovGlick1965,KitagawaUeda1993}:
\begin{equation}
H(t)=\Omega_0 S_z+\frac{\kappa}{N_s}S_z^2+A\cos(\omega t)\,S_x,
\label{eq:ex3_collective_H}
\end{equation}
where $S_\mu\equiv \frac12\sum_{j=1}^{N_s}\sigma_j^\mu,\ \mu=x,y,z$.
The $S_z^2$ term corresponds to all-to-all Ising couplings in the microscopic spin language and renders the model genuinely interacting.
Because $[H(t),S^2]=0$ for all $t$, dynamics initiated in the fully symmetric subspace (total spin $S=N_s/2$) remains in that subspace, whose Hilbert-space dimension is
$N=2S+1=N_s+1$.
We focus on adiabatic following of the instantaneous ground state $\ket{E_0(t)}$.

\begin{figure}[t]
\centering
\includegraphics[width=\columnwidth]{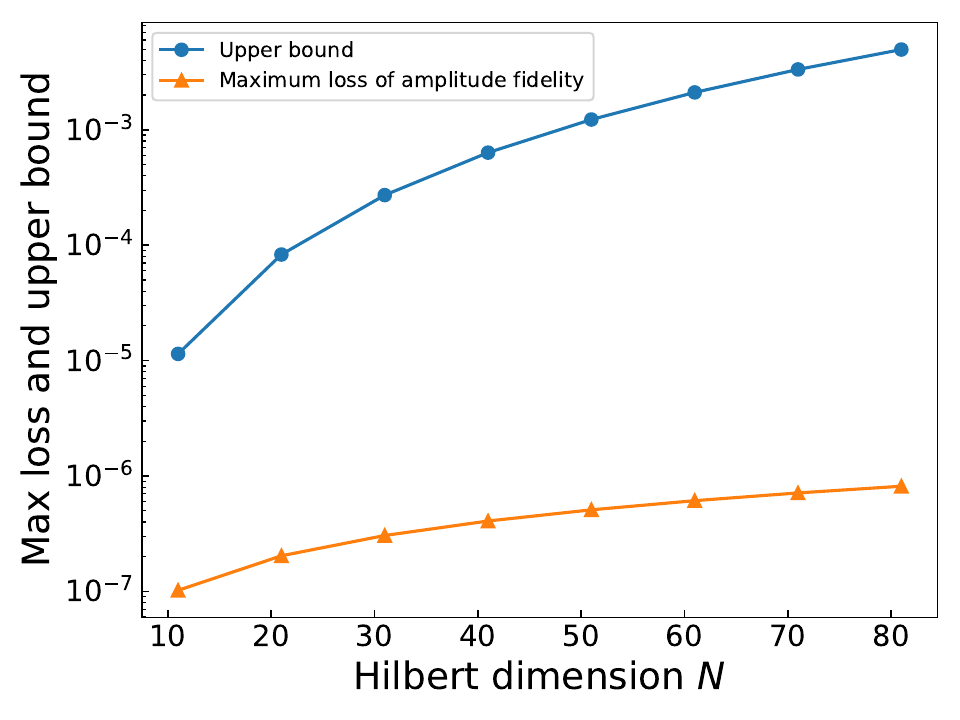}
\caption{Mitigation of the  $N$-scaling issue in the geometric Floquet condition.
Parameters are $\Omega_0=15$, $\kappa=1$, $A=0.002$, and $\omega=\Omega_0(N_s+1)$.}
\label{fig:scaling_absent}
\end{figure}

For each $N_s$ we extract from one driving cycle the two stroboscopic quantities entering Theorem~\ref{thm:main}:
the one-period Fubini--Study length $\mathcal{L}_0$ of the instantaneous ground-state ray and the quasienergy-separation measure $g$.
The circles in Fig.~\ref{fig:scaling_absent} show the corresponding right-hand side of the time-uniform fidelity-amplitude bound,
which carries no explicit prefactor that grows with $N$.
The triangles show the numerically observed worst-case fidelity-amplitude loss over $50$ driving periods,
$\max_t\!\left(1-\abs{d_0(t)}\right)$.
Notably, the geometric Floquet bound remains below $10^{-2}$ up to $N=81$.
This confirms, in a genuinely interacting many-body setting, that the geometric Floquet condition~\eqref{eq:cond_main} mitigates the common ``scaling issue'' that can arise when adiabatic criteria are formulated with state-by-state couplings and loose sums over an extensive number of levels. In larger spectra, the state-targeted version can further reduce conservativeness by ignoring near-degeneracies that do not involve the dominant Floquet component of the tracked state.


\section{Discussion and outlook}
Adiabaticity is often equated with slow or low-frequency driving.
The examples above show instead that, in strictly periodic systems, the decisive issue is the combination of one-cycle geometric rotation and distance from quasienergy resonances.
This suggests a use case for fast periodic protocols that remain adiabatic under indefinite repetition, for instance in adiabatic strokes of quantum heat engines \cite{Quan2007} or in calibration routines where repeated application of a control waveform should not populate unwanted states.

The geometric Floquet condition should be viewed as complementary to shortcuts to adiabaticity, numerical optimal control, and broader Floquet-engineering strategies.
Those frameworks are primarily tools for {designing} controls; our theorem is a tool for {certifying} a fixed periodic control.
Its payoff is a rigorous time-uniform guarantee based on one-cycle data.
Its limitations are equally clear: the system is assumed closed, finite-dimensional, Hermitian, and exactly periodic, and the tracked instantaneous eigenray must remain smooth and nondegenerate.
The criterion is sufficient rather than necessary, so there are adiabatic regimes it does not certify.
The state-targeted refinement introduced above partly mitigates this conservativeness when only a specified branch is relevant.

The absence of an explicit $N$-prefactor in the geometric Fubini--Study length removes one common source of unfavorable scaling.
However, the scaling of the relevant quasienergy gap with system size remains model- and regime-dependent, and a systematic characterization of that behavior is an important direction for future work.
Likewise, while small $g$ flags parameter regions where Floquet resonances can enable nonadiabatic transitions and related phenomena \cite{Russomanno2012,Shu2018,Claeys2018,TapiasArze2020,Mishra2021}, violation of Eq.~\eqref{eq:cond_main} alone does not guarantee such effects.
Extending the present framework to noisy, open, non-Hermitian, or slowly modulated periodic settings, and quantifying robustness against experimental estimation error in $U(t_0+T,t_0)$, are important next steps.

\begin{acknowledgments}
This work was supported as part of the Center for Molecular Magnetic Quantum Materials, an Energy Frontier Research Center funded by the U.S. Department of Energy, Office of Science, Basic Energy Sciences under Award no. DE-SC0019330.
\end{acknowledgments}


\appendix

\section*{Appendix}

This appendix provides derivations of Theorem~\ref{thm:main}, its corollary, and the state-targeted refinement quoted in the main text.

\subsection{A geometric bound in the instantaneous eigenbasis}
We first bound the overlap with an instantaneous eigenstate in terms of the Fubini--Study length traversed by that eigenray.

\begin{lemma}[Fubini--Study bound]
\label{lem:FS}
Let $\ket{\Psi(t)}$ solve $\ii\partial_t\ket{\Psi(t)}=H(t)\ket{\Psi(t)}$ with $\ket{\Psi(t_0)}=\ket{E_n(t_0)}$.
Define $d_n(t)=\braket{E_n(t)}{\Psi(t)}$ and $\mathcal{L}_n(\tau)=\int_{t_0}^{t_0+\tau} v_n(t)\,\mathrm{d}t$ with $v_n$ as in~\eqref{eq:FS_speed}.
If $\mathcal{L}_n(\tau) \le \pi/2$, then for any $\tau\ge 0$,
\begin{equation}
\label{eq:lem_FS_result}
|d_n(t_0+\tau)| \ge \cos\!\bigl(\mathcal{L}_n(\tau)\bigr)
= 1-2\sin^2\!\left(\frac{\mathcal{L}_n(\tau)}{2}\right).
\end{equation}
\end{lemma}

Proof: Expand $\ket{\Psi(t)}$ in the instantaneous eigenbasis with dynamical and geometric phases removed:
\[
\ket{\Psi(t)}=\sum_m b_m(t)\,e^{-\ii\int_{t_0}^t E_m(t')\mathrm{d}t'}\,e^{-\int_{t_0}^t \langle E_m|\dot E_m\rangle\, \mathrm{d}t '}\,\ket{E_m(t)},
\]
so that $b_m(t_0)=\delta_{mn}$ and the coefficients satisfy
\[
\begin{aligned}
\dot b_n(t) = -\sum_{m\neq n} b_m(t)\,\langle E_n(t)|\dot E_m(t)\rangle\,e^{-\ii\int_{t_0}^t (E_m-E_n)\, \mathrm{d}t '}\\ \times e^{-\int_{t_0}^t (\langle E_m|\dot E_m\rangle-\langle E_n|\dot E_n\rangle)\, \mathrm{d}t '}.	
\end{aligned}
\]
Taking absolute values and using Cauchy--Schwarz gives
\[
\begin{aligned}
\abs{\dot b_n(t)}
&\le \sum_{m\neq n}\abs{b_m(t)}\,\abs{\langle E_n(t)|\dot E_m(t)\rangle}\\
&\le \left(\sum_{m\neq n}\abs{b_m(t)}^2\right)^{1/2}
\left(\sum_{m\neq n}\abs{\langle E_n(t)|\dot E_m(t)\rangle}^2\right)^{1/2}\nonumber\\
&= \sqrt{1-\abs{b_n(t)}^2}\;\,v_n(t).
\end{aligned}
\]
Let $u(t)\equiv \abs{b_n(t)}\in[0,1]$. Where differentiable,
\[
\frac{\mathrm{d}}{ \mathrm{d}t }\arccos u(t) = -\frac{\dot u(t)}{\sqrt{1-u(t)^2}} \le v_n(t),
\]
since $\dot u\ge -\abs{\dot b_n}$.
Integrating from $t_0$ to $t_0+\tau$ and using $u(t_0)=1$ yields
\[
\arccos u(t_0+\tau)\le \int_{t_0}^{t_0+\tau} v_n(t)\, \mathrm{d}t  = \mathcal{L}_n(\tau),
\]
hence $u(t_0+\tau)\ge \cos(\mathcal{L}_n(\tau))$ if $\mathcal{L}_n(\tau) \le \pi/2$.
Because $|d_n(t)|=|b_n(t)|$ in this gauge, Eq.~\eqref{eq:lem_FS_result} follows.

Specializing Lemma~\ref{lem:FS} to $\tau=T$ gives
\begin{equation}
\label{eq:1minusd_bound}
1-|d_n(t_0+T)| \le 2\sin^2\!\left(\frac{\mathcal{L}_n}{2}\right),
\qquad \mathcal{L}_n \equiv \mathcal{L}_n(T).
\end{equation}

\subsection{A stroboscopic Floquet mixing bound}
Assume the periodic gauge~\eqref{eq:periodic_gauge}.
Using Floquet expansion at $t_0$,
\[
\ket{\Psi(t_0+T)}=\sum_\alpha a_\alpha\,e^{-\ii\epsilon_\alpha T}\,\ket{\phi_\alpha(t_0)},
\]
and $\ket{E_n(t_0+T)}=\ket{E_n(t_0)}=\sum_\alpha a_\alpha \ket{\phi_\alpha(t_0)}$.
Therefore
\[
d_n(t_0+T)= \sum_\alpha |a_\alpha|^2 e^{-\ii\epsilon_\alpha T}=\sum_\alpha c_\alpha e^{-\ii\epsilon_\alpha T}.
\]
Compute the squared modulus:
\[
\begin{aligned}
|d_n(t_0+T)|^2
&=\sum_\alpha c_\alpha^2 + 2\sum_{\alpha<\beta} c_\alpha c_\beta \cos\!\bigl((\epsilon_\alpha-\epsilon_\beta)T\bigr)\nonumber\\
&=1-4\sum_{\alpha<\beta} c_\alpha c_\beta \sin^2\!\left(\frac{(\epsilon_\alpha-\epsilon_\beta)T}{2}\right).
\end{aligned}
\]
Let
\begin{equation}
\label{eq:S_def}
S \equiv \sum_{\alpha<\beta} c_\alpha c_\beta \sin^2\!\left(\frac{(\epsilon_\alpha-\epsilon_\beta)T}{2}\right)\in\left[0,\frac14\right].
\end{equation}
Then $|d|=\sqrt{1-4S}\le 1-2S$ (since $\sqrt{1-x}\le 1-x/2$ for $x\in[0,1]$), hence
\begin{equation}
\label{eq:S_vs_d}
1-|d_n(t_0+T)| \ge 2S.
\end{equation}
Using the definition of $g$ in~\eqref{eq:g_def}, we have
\begin{equation}
\label{eq:S_ge_g2_pairsum}
S \ge g^2 \sum_{\alpha<\beta} c_\alpha c_\beta,
\end{equation}
and combining with~\eqref{eq:S_vs_d} gives
\begin{equation}
\label{eq:pair_sum_bound_from_d}
\sum_{\alpha<\beta} c_\alpha c_\beta \le \frac{1-|d_n(t_0+T)|}{2g^2}.
\end{equation}
Finally insert~\eqref{eq:1minusd_bound}:
\begin{equation}
\label{eq:delta_def}
\sum_{\alpha<\beta} c_\alpha c_\beta \le \delta,
\qquad
\delta \equiv \frac{\sin^2(\mathcal{L}_n/2)}{g^2}.
\end{equation}
 \subsection{Dominant Floquet component and uniform-in-time fidelity}
Let $\alpha_0$ be an index maximizing $c_\alpha$, i.e.,
$c_{\alpha_0}=\max_\alpha c_\alpha .
$ From Eq.~\eqref{eq:delta_def} we have
$\sum_{\alpha<\beta} c_\alpha c_\beta \le \delta .
$ Using $\big(\sum_\alpha c_\alpha\big)^2=\sum_\alpha c_\alpha^2+2\sum_{\alpha<\beta}c_\alpha c_\beta$ and $\sum_\alpha c_\alpha=1$, this is equivalent to
$1-\sum_\alpha c_\alpha^2
=2\sum_{\alpha<\beta} c_\alpha c_\beta
\le 2\delta,
$ hence $\sum_\alpha c_\alpha^2 \ge 1-2\delta$.
Moreover, since $\sum_\alpha c_\alpha^2 \le c_{\alpha_0}\sum_\alpha c_\alpha=c_{\alpha_0}$, we obtain
\begin{equation}
\label{eq:c_alpha0_lower}
c_{\alpha_0} \ge 1-2\delta .
\end{equation}
When $\delta<1/4$, we have $c_{\alpha_0}>1/2$, so the maximizing index $\alpha_0$ is unique.

Now restart the same one-period argument at an arbitrary time $t$ with initial state $\ket{E_n(t)}$.
This yields an index $\tilde\alpha(t)$ such that
\begin{equation}
\label{eq:En_close_to_floquet}
\abs{\braket{\phi_{\tilde\alpha(t)}(t)}{E_n(t)}}^2 \ge 1-2\delta.
\end{equation}
When $\delta<1/4$, the index $\tilde\alpha(t)$ is unique because $1-2\delta>1/2$ and
$\sum_\alpha \abs{\braket{\phi_{\alpha}(t)}{E_n(t)}}^2=1$.
Since each overlap $\abs{\braket{\phi_{\alpha}(t)}{E_n(t)}}^2$ depends continuously on $t$, the discrete unique index $\tilde\alpha(t)$ cannot change with $t$.
Thus $\tilde\alpha(t)=\tilde\alpha(t_0)=\alpha_0$ for all $t$.

On the other hand, in the Floquet decomposition of the evolving state,
$\braket{\phi_{\alpha}(t)}{\Psi(t)}=a_\alpha e^{-i\epsilon_\alpha(t-t_0)}$, so
$\abs{\braket{\phi_{\alpha}(t)}{\Psi(t)}}^2=|a_\alpha|^2\equiv c_\alpha$ is independent of $t$.
Therefore,
\begin{equation}
\label{eq:phi_alpha0Psi_overlap}
\abs{\braket{\phi_{\alpha_0}(t)}{\Psi(t)}}^2 = |a_{\alpha_0}|^2=c_{\alpha_0}\ge 1-2\delta.
\end{equation}

Now bound the desired overlap using the decomposition along $\ket{\phi_{\alpha_0}(t)}$ and its orthogonal complement:
\begin{equation}
\label{eq:EnPsi_overlap_bound}
\begin{aligned}
&\abs{\braket{E_n(t)}{\Psi(t)}}
\ge
\abs{\braket{E_n(t)}{\phi_{\alpha_0}(t)}}\abs{\braket{\phi_{\alpha_0}(t)}{\Psi(t)}} \\
 &-\sqrt{1-\abs{\braket{E_n(t)}{\phi_{\alpha_0}(t)}}^2}\,
 \sqrt{1-\abs{\braket{\phi_{\alpha_0}(t)}{\Psi(t)}}^2} \\
&\qquad \qquad \qquad  \ge (1-2\delta) - 2\delta
=1-4\delta,
\end{aligned}
\end{equation}
where we have used the triangle inequality and the Cauchy--Schwarz inequality.
Therefore, if $\delta \le \varepsilon/4$, then $\abs{\braket{E_n(t)}{\Psi(t)}}\ge 1-\varepsilon$ for all $t\ge t_0$.
This completes the proof of Theorem~\ref{thm:main}.

If $\varepsilon\ll 1$ (the practically relevant regime), $\mathcal{L}_n \le \pi/4$ is satisfied automatically.
Using $|\sin x| \le |x|$ once again, Theorem~\ref{thm:main} leads to the geometric Floquet condition~\eqref{eq:cond_main}.

\subsection{State-targeted variant}
The appearance of the global (worst-case) gap $g$ is tied to a single step: in going from Eq.~\eqref{eq:S_def} to Eq.~\eqref{eq:S_ge_g2_pairsum} we lower-bound {every} factor $\sin^2[(\epsilon_\alpha-\epsilon_\beta)T/2]$ by the same minimum $g^2$. A mode-resolved (state-targeted) refinement follows by fixing the Floquet label $\alpha_0$ that is relevant to the chosen initial eigenstate (e.g., the dominant component identified in the previous subsection) and defining the {targeted} stroboscopic gap [Eq.~\eqref{eq:g_targeted_def}]
\[
g_{\alpha_0}\equiv \min_{\beta\neq \alpha_0}\left|\sin\!\left(\frac{(\epsilon_{\alpha_0}-\epsilon_\beta)T}{2}\right)\right|.
\]
Then Eq.~\eqref{eq:S_ge_g2_pairsum} is replaced by the restricted bound
\[
S\;\ge\;\sum_{\beta\neq\alpha_0}c_{\alpha_0}c_\beta\,
\sin^2\!\left(\frac{(\epsilon_{\alpha_0}-\epsilon_\beta)T}{2}\right)
\;\ge\;g_{\alpha_0}^2\,c_{\alpha_0}(1-c_{\alpha_0}),
\]
so Eqs.~\eqref{eq:pair_sum_bound_from_d} and \eqref{eq:delta_def} become
$c_{\alpha_0}(1-c_{\alpha_0})\le [1-|d_n(t_0+T)|]/(2g_{\alpha_0}^2)\le
\sin^2(\mathcal{L}_n/2)/g_{\alpha_0}^2\equiv \delta_{\alpha_0}$.
Downstream, the only bookkeeping change is that the global pair-sum estimate leading to Eq.~\eqref{eq:c_alpha0_lower} is replaced by the quadratic constraint
$c_{\alpha_0}(1-c_{\alpha_0})\le\delta_{\alpha_0}$ (which, on the large-overlap branch, implies $c_{\alpha_0}\ge (1+\sqrt{1-4\delta_{\alpha_0}})/2$, hence $c_{\alpha_0}\ge 1-2\delta_{\alpha_0}$ for $\delta_{\alpha_0}<1/4$), and in the restart step one applies the same targeted inequality to the {same} index $\alpha_0$ (so Eq.~\eqref{eq:En_close_to_floquet} is replaced by
$|\langle\phi_{\alpha_0}(t)|E_n(t)\rangle|^2\ge 1-2\delta_{\alpha_0}$, with continuity preventing a branch switch). With these substitutions, Eqs.~\eqref{eq:phi_alpha0Psi_overlap} and~\eqref{eq:EnPsi_overlap_bound} proceed as written and yield the same uniform-in-time bound as Theorem~\ref{thm:main} with $g$ replaced by $g_{\alpha_0}$.

\bibliography{ref_abbr}

\end{document}